\documentclass[pra,twocolumn,amsmath,amssymb]{revtex4-1} %

\usepackage{epsfig,amsmath}
\usepackage{subfigure}
\usepackage{graphicx}% Include figure files
\usepackage{dcolumn}% Align table columns on decimal point
\usepackage{stmaryrd}
\usepackage{mathrsfs}
\usepackage{pifont}
\usepackage{amsthm}
\usepackage{amssymb}
\usepackage{bm}
\usepackage{latexsym}
\usepackage[colorlinks=true,linkcolor=blue,citecolor=blue]{hyperref}
\usepackage{color}
\usepackage{soul}
\theoremstyle{plain}

\newcommand{\la}{\langle}
\newcommand{\ra}{\rangle}

\newcommand{\ti}{\tilde}
\newcommand{\ga}{\gamma}
\newcommand{\Ga}{\Gamma}

\newcommand{\De}{\Delta}

\newcommand{\om}{\omega}
\newcommand{\Om}{\Omega}
\newcommand{\de}{\delta}

\newcommand{\non}{\nonumber}
\newcommand{\pa}{\partial}

\def\pra#1{{ Phys.\ Rev. A\/} {\bf#1}}
\def\prb#1{{ Phys.\ Rev. B\/} {\bf#1}}

\def\prl#1{{ Phys.\ Rev.\ Lett.} {\bf#1}}

\def\sci#1{{ Science} {\bf#1}}

\def\pla#1{{ Phys.\ Lett. A\/} {\bf#1}}
\def\rmp#1{{ Rev. \ Mod. \ Phys.} {\bf#1}}
\def\nat#1{{ Nature} {\bf#1}}

\def\njp#1{{ New. J. \ Phys.} {\bf#1}}

\begin{document}

\title{Non-Abelian holonomic transformation in the presence of classical noise}

\author{Jun Jing$^{1,2,3,4}$, Chi-Hang Lam$^{2}$, and Lian-Ao Wu$^{3,5}$}\thanks{Author to whom any correspondence should be addressed. Email address: lianao.wu@ehu.es}

\affiliation{$^{1}$Institute of Atomic and Molecular Physics and Jilin Provincial Key Laboratory of Applied Atomic and Molecular Spectroscopy, Jilin University, Changchun 130012, Jilin, China \\ $^{2}$Department of Applied Physics, Hong Kong Polytechnic University, Hung Hom, Hong Kong, China \\ $^{3}$Department of Theoretical Physics and History of Science, The Basque Country University (EHU/UPV), PO Box 644, 48080 Bilbao, Spain \\ $^{4}$Department of Physics, Zhejiang University, Hangzhou 310027, Zhejiang, China \\ $^{5}$Ikerbasque, Basque Foundation for Science, 48011 Bilbao, Spain}

\date{\today}

\begin{abstract}
It is proposed that high-speed universal quantum gates can be realized by using non-Abelian holonomic transformation. A cyclic evolution path which brings the system periodically back to a degenerate qubit subspace is crucial to holonomic quantum computing. The cyclic nature and the resulting gate operations are fully dependent on the precise control of driving parameters, such as the modulated envelop function of Rabi frequency and the control phases. We investigate the effects of fluctuations in these driving parameters on the transformation fidelity of a universal set of single-qubit quantum gates. We compare the damage effects from different noise sources and determine the ``sweet spots'' in the driving parameter space. The nonadiabatic non-Abelian quantum gate is found to be more susceptible to classical noises on the envelop function than that on the control phases. We also extend our study to a two-qubit quantum gate.
\end{abstract}


\maketitle

\section{Introduction}

Quantum geometrical phase~\cite{Berry,Wilczek,Aharonov,Anandan}, which is proportional to the area spanned in the parameter space but is insensitive to the trajectory followed by the system, has inspired more and more efforts in circuit-based quantum computation and quantum control protocols~\cite{Deutsch,Siu,Mizel}. An appealing modern application of quantum geometrical phase is non-Abelian holonomic quantum computation (HQC)~\cite{AnandanPLA,Zanardi,Vedral,Duan,Salomaa,Wu,Lidar,nanaholonomic}, in which one performs a universal set of unitary transformations, i.e., quantum logic gates operations via cyclic evolution in a degenerate subspace. More generally, HQC belongs to the field of quantum state engineering~\cite{Marcelo,Adolfo}. It is argued that the implementation of quantum gates encoded in a degenerate subspace suppresses the effect of dynamical phase around the given loops in the parameter space. Thus it is not surprising to see that the conventional HQC schemes are based on adiabatic evolution due to its resilience against local fluctuations. The adiabatic theorem~\cite{Born,Messiah} asserts that at any moment a quantum system remains closely at an instantaneous eigenstate of a slowly varying Hamiltonian. Specifically for a cyclic adiabatic process, a geometric phase is acquired over the course of the cycle~\cite{Bergmann,Zhu,Kral,Hu}. Experimental implementations of adiabatic HQC have been proposed in various physical systems, such as trapped ions~\cite{Duan}, superconducting nanocircuits~\cite{Falci}, semiconductor quantum dots~\cite{Solinas}, to name a few.

Despite the advantages such as robustness to fluctuations in runtime and system energy, geometric operations in adiabatic HQC suffer from a dilemma between a long runtime and a good coherence, noting that a loss of coherence can occur due to fluctuations in the control parameters. Nonadiabatic non-Abelian geometric-phase-based holonomic transformation~\cite{nanaholonomic,xu,robustnonahg,long,error0,DuanNat,Zanardi2,Yan,Villar} has been recently proposed to demonstrate universal operations for quantum computation in both theory and experiment. As an all-geometric scheme, it still retains the advantages of the conventional HQC but the evolution speed can be greatly accelerated. There have been studies on the reliability of the nonadiabatic non-Abelian quantum gate upon considerations of adverse effects including gate decoherence and noise~\cite{error1}, influence from the Lindbladian~\cite{error2}, systematical error~\cite{error3,error4}, rotating-wave approximation~\cite{error0} and finite operational time~\cite{error5}. The robustness of HQC, in particular the performance of the unitary transformation over general input states, against classical noise (fluctuations) in the control Hamiltonian parameters is still under investigation~\cite{error4}. And that would be the focus in this work.

Classical noise characterizing small system perturbations can have a dramatic impact on the cyclic time evolution of system as well as the performance of quantum gates which require precisely controlled external driving. In this work, we introduce classical noises in the form of stochastic fluctuations in the control parameters~\cite{noise,Inverse} of a driven Hamiltonian of a three-level atom or ion forming a $\Lambda$-configuration for realizing universal holonomic single-qubit and two-qubit gates. Such fluctuations are often due to imprecise system controls and other unknown environmental influences. They can be introduced during gate operations or during reading of the results. Then, the output (resulting) quantum state under the operation by the perturbed holonomic quantum gate will in general deviate from that under the noise-free unitary transformation. We study a transformation fidelity for a general input state, which quantitatively measures these deviations. In particular, we estimate the robustness of the nonadiabatic holonomic transformation and determine the conditions in which the desired state passage can be reliably realized in the presence of classical noise. We note that both the magnitude and the correlation of the classical noises are important factors in determining the transformation fidelity.

The rest of this work is organized as follows. In Sec.~\ref{theory}, we introduce a universal nonadiabatic non-Abelian quantum gate implementation. The logical subspace and control parameters including the envelop function and two control phases are explained. In Sec.~\ref{random}, we consider stochastic fluctuations in each parameter and their damage to the fidelity of the nonadiabatic holonomic transformation. We analyze the fidelity as a function of or minimized over the input (initial) states and the driving parameters. In Sec~\ref{discuss}, we extend our formalism to a two-qubit gate case. A conclusion is presented in Sec.~\ref{conc}.

\section{Constructing nonadiabatic holonomic quantum gates}\label{theory}

We first construct a universal holonomic single-qubit gate based on a driven $\Lambda$-configuration three-level system as well as nonadiabatic non-Abelian geometric transformations. The driving Hamiltonian, realized by system-laser interactions, admits classical noise originated from the control lasers and environmental disturbance.

The bare three-level system consists of two nondegenerate ground states $|0\ra$ and $|1\ra$, representing the logic states in the encoded quantum gate, and one excited state $|e\ra$ acting as the auxiliary state. In the presence of two separable polarized laser pulses properly tuned to be at resonance with transitions $|e\ra\leftrightarrow|1\ra$ and $|e\ra\leftrightarrow|0\ra$, respectively. Assuming the level $|0\ra$ has an energy $\om_0=0$, without loss of generality, the Hamiltonian can be written as
\begin{eqnarray}\non
H_0&=&\om_e|e\ra\la e|+\om_1|1\ra\la 1|\\ \label{H0} &+&\Om(t)[a_0(t)|e\ra\la0|+b_0(t)|e\ra\la1|+h.c.],
\end{eqnarray}
where $\om_e$ and $\om_1$ are the bare energies of $|e\ra$ and $|1\ra$, respectively, $\Om(t)$ is the modulated Rabi frequency (pulse envelop or amplitude), and $a_0(t)$ and $b_0(t)$ are the driving coefficients assumed to satisfy $|a_0(t)|^2+|b_0(t)|^2=1$ for simplicity. To cancel the bare energy terms in the original Hamiltonian~(\ref{H0}), we turn to the rotating frame by applying the unitary transformation $U_0=\exp[i(\om_e|e\ra\la e|+\om_1|1\ra\la 1|)t]$. Upon this rotation, the Hamiltonian becomes
\begin{equation}\label{H1}
H_1(t)=\Om(t)[a(t)|e\ra\la0|+b(t)|e\ra\la1|+h.c.].
\end{equation}
Here the coefficients are
\begin{equation*}
 a(t)=a_0(t)e^{i\om_et}, \quad b(t)=b_0(t)e^{i(\om_e-\om_1)t},
\end{equation*}
which still satisfy the normalization condition $|a(t)|^2+|b(t)|^2=1$. Their time dependence can be suppressed by choosing $a_0(t)\propto e^{-i\om_et}$ and $b_0(t)\propto e^{-i(\om_e-\om_1)t}$. In general, the {\it time-independent} coefficients $a$ and $b$ can further be parameterized by two control phases in the form
\begin{equation*}
a=\sin\frac{\theta}{2}e^{i\phi}, \quad b=\cos\frac{\theta}{2}.
\end{equation*}
Then, we have three parameters $\Om(t)$, $\theta$, and $\phi$ controllable via the two driving lasers. They are taken as real numbers in the ideal case of stable control. In the following, we will show that the envelop function of the Rabi frequency $\Om(t)$ determines the cyclic period as well as the speed of the quantum gate operation, while the control phases $\theta$ and $\phi$ specify the type of the quantum gate.

A spectral analysis of the Hamiltonian in Eq.~(\ref{H1}) gives
\begin{equation}
H_1(t)=\Om(t)(|\psi_{b+}\ra\la\psi_{b+}|-|\psi_{b-}\ra\la\psi_{b-}|)
+0|\psi_d\ra\la\psi_d|.
\end{equation}
In terms of the basis states $\{|0\ra, |1\ra, |e\ra\}$, $|\psi_{b\pm}\ra=(1/\sqrt{2})[a, b, \pm1]'$ represent two bright eigenstates, while $|\psi_d\ra=[b, -a, 0]'$ is a state playing no role in the dynamics and the gate operation. The time-evolution operator resulting from $H_1$ is found to be
\begin{widetext}
\begin{equation}\label{U1}
U(t)=e^{-i\int_0^tdsH_1(s)}=
\left(\begin{array}{ccc}\sin^2\frac{\theta}{2}\cos\bar{\Om}+\cos^2\frac{\theta}{2} & \frac{\sin\theta}{2}e^{-i\phi}(\cos\bar{\Om}-1) & i\sin\frac{\theta}{2}e^{-i\phi}\sin\bar{\Om}\\ \frac{\sin\theta}{2}e^{i\phi}(\cos\bar{\Om}-1) & \cos^2\frac{\theta}{2}\cos\bar{\Om}+\sin^2\frac{\theta}{2} & i\cos\frac{\theta}{2}\sin\bar{\Om} \\  i\sin\frac{\theta}{2}e^{i\phi}\sin\bar{\Om} & i\cos\frac{\theta}{2}\sin\bar{\Om} & \cos\bar{\Om}\end{array}\right),
\end{equation}
\end{widetext}
where $\bar{\Om}\equiv\bar{\Om}(t)=\int_0^tds\Om(s)$. It is straightforward to see that when $\bar{\Om}(T)=\pi$, i.e., $\int_0^Tdt\Om(t)=\pi$, the first two degrees of freedom, $|0\ra$ and $|1\ra$, will be decoupled from the excited (ancillary) state $|e\ra$. It follows that the qubit space spanned by $|0\ra$ and $|1\ra$ is invariant under the time evolution $U(s)$ if the lasers satisfy $\bar{\Om}(T)=\pi$. It can be verified that this evolution is purely geometric since $\la k| U^\dag(s) H_1(s)U(s) |l\ra=\la k|H_1(s)|l\ra=0$ for $k,l=0,1$ and $s\in[0, t]$.

Under the above conditions, the final time evolution operator $U(T)$ is projected onto the qubit subspace spanned by $|0\ra$ and $|1\ra$ and can be expressed as
\begin{equation}\label{Uh}
U_h(T)=\left(\begin{array}{cc} \cos\theta & -\sin\theta e^{-i\phi} \\ -\sin\theta e^{i\phi} & -\cos\theta \end{array}\right).
\end{equation}
It can be used to realize any single-qubit rotation, i.e., an arbitrary unitary transformation for a single qubit. This thus defines a universal holonomic single-qubit gate. For examples, Eq.~(\ref{Uh}) can realize (i) the Hadamard gate with $\theta=\frac{3\pi}{4}$ and $\phi=0$, (ii) the Pauli-X gate with $\theta=\frac{\pi}{2}$ and $\phi=\pi$, (iii) the Pauli-Z gate with $\theta=0$, and (iv) the phase-shift gate with $\theta=\frac{3\pi}{2}$. In general, using Eq.~(\ref{Uh}), an input (initial) state $|\psi(0)\ra=\alpha e^{i\eta}|0\ra+\beta|1\ra$ will be transformed into
\begin{eqnarray}\non
|\Psi(T)\ra&=&(\alpha\cos\theta e^{i\eta}-\beta\sin\theta e^{-i\phi})|0\ra\\ \label{PsiT} &-&(\alpha\sin\theta e^{i(\eta+\phi)}+\beta\cos\theta)|1\ra,
\end{eqnarray}
where $\alpha$, $\beta$ and $\eta$ are assumed to be real numbers satisfying $\alpha^2+\beta^2=1$. Here and in the following, $\Psi$ denotes the resulting state from an ideal noise-free holonomic transformation.

The gate operation provided by Eq.~(\ref{Uh}) is in general universal connecting any pair of pure states and there is in principle no limit on the operation time $T$. The control parameters $\theta$ and $\phi$ set up the desired quantum gate. The effective time evolution operator $U_h$ in Eq.~(\ref{Uh}) thus provides a general protocol for quantum state engineering. It is therefore important to consider the reliability of this gate operation.

\section{Reliability of holonomic transformation in the presence of classical noise}\label{random}

The ideal holonomic transformation specified by Eq.~(\ref{PsiT}) is not always attainable once nonideal driving in the original Hamiltonian~(\ref{H0}) is taken into account. We now consider the stochastic time-evolution operator $U_\xi(t)$ which deviates from $U(t)$ in Eq.~(\ref{U1}) under the effect of a single noisy control parameter $\xi\in\{\Om, \theta, \phi\}$. To measure the robustness of the holonomic quantum gate for HQC, we study a transformation fidelity defined by
\begin{equation}\label{fidelity}
\mathcal{F}_\xi=M[\la\Psi(T)|\psi_\xi(T)\ra\la\psi_\xi(T)|\Psi(T)\ra].
\end{equation}
Here $M[\cdot]$ means the ensemble average over all random realizations of fluctuations in the control parameter, and $|\psi_\xi(t)\ra\equiv U_\xi(t)|\psi_0\ra$ denotes the nonideal output state of the noisy quantum gate. The transformation fidelity indicates the leakage of the output state out of the logic subspace. For example, putting $\xi\equiv\Om$, $U_{\Om}(t)$ can be obtained after letting $\Om(t)\rightarrow\Om'(t)=\Om(t)+\de_\Om(t)$ in Eq.~(\ref{U1}). It yields
\begin{eqnarray*}
|\psi_\Om(t)\ra&=&\bigg[\cos\bar{\Om}'(\alpha\sin^2\frac{\theta}{2}e^{i\eta}
+\beta\frac{\sin\theta}{2}e^{-i\phi}) \\&+&\alpha\cos^2\frac{\theta}{2}e^{i\eta}
-\beta\frac{\sin\theta}{2}e^{-i\phi}\bigg]|0\ra\\ &+&
\bigg[\cos\bar{\Om}'(\beta\cos^2\frac{\theta}{2}
+\alpha\frac{\sin\theta}{2}e^{i(\phi+\eta)}) \\&+& \beta\sin^2\frac{\theta}{2}
-\alpha\frac{\sin\theta}{2}e^{i(\phi+\eta)}\bigg]|1\ra \\ &+&
i\sin\bar{\Om}'(\beta\cos\frac{\theta}{2}+\alpha\sin\frac{\theta}{2}e^{i(\phi+\eta)})|e\ra,
\end{eqnarray*}
where $\bar{\Om}'=\bar{\Om}+\int_0^tds\de_\Om(s)$. It can be verified (see, e.g., Ref.~\cite{noise}) that for any classical Gaussian noise $\de_\xi(t)$ with a zero mean $\la\de_\xi(t)\ra=0$ and an auto-correlation function $C_\xi(t,s) = \la\de_\xi(t)\de_\xi(s)\ra$,
\begin{eqnarray*}
M\left[e^{im\int_0^tdt_1\de_\xi(t_1)}\right]&=&
e^{-m^2\int_0^tdt_1\int_0^{t_1}dt_2C_\xi(t_1,t_2)}, \\
M\left[e^{im\de_\xi(t)}\right]&=&e^{-\frac{m^2}{2}C_\xi(t,t)},
\end{eqnarray*}
where $m$ is a real constant. These results are helpful in evaluation of the fidelity in Eq.~(\ref{fidelity}).

The classical noise sources in our work are associated with the three driving parameters, i.e., the amplitude-envelop function $\Om$ and two phases $\theta$ and $\phi$ in Rabi frequency. The shape and duration of the input laser field are determined by the envelop function $\Om$, whereas the carrier-envelop phases of the two driving lasers are described by $\theta$ and $\phi$. Physically the amplitude and the phases can be tuned by certain combinations of acousto-optical modulators and phase-modulation locking achieved with low driving voltages, respectively. Recently, techniques to separately modulate the amplitude and phases of laser sources have been proposed and experimentally demonstrated~\cite{exp1,exp2,exp3,exp4}. In the following, we assume that only one control parameter admits significant fluctuations in each case and the fluctuations are of Gaussian type. However, the conclusion is independent of the spectral function or the correlation function of the noise.

\subsection{Fidelity under noisy envelop function $\Om$}

The fidelity in the presence of fluctuations in $\Om(t)$ can be obtained from the overlap between the ideal and nonideal wave functions,
\begin{equation*}
\la\Psi(T)|\psi_\Om(T)\ra=[1+\cos\bar{\Om}'(T)]f(\theta,\phi)-\cos\bar{\Om}'(T),
\end{equation*}
where
\begin{eqnarray*}
f=f(\theta,\phi)&\equiv&\alpha^2\cos^2\frac{\theta}{2}+\beta^2\sin^2\frac{\theta}{2}
-\alpha\beta\sin\theta\cos(\phi+\eta).
\end{eqnarray*}
Note that the tight upper and lower bounds of $f(\theta,\phi)$ are limited by $|\alpha\cos(\theta/2)-\beta\sin(\theta/2)|^2$ or $|\alpha\cos(\theta/2)+\beta\sin(\theta/2)|^2$. Either situation satisfies $0\leq f\leq1$. Substituting the above results into Eq.~(\ref{fidelity}) with $\bar{\Om}(T)=\pi$ and performing the ensemble average for the fluctuations, the fidelity is obtained as
\begin{eqnarray}\non
\mathcal{F}_{\Om}&=&\frac{1+e^{-4\bar{C}_{\Om}(T)}}{2}+
\left[2e^{-\bar{C}_{\Om}(T)}-e^{-4\bar{C}_{\Om}(T)}-1\right]f \\ \label{fidOm} &+&\frac{3-4e^{-\bar{C}_{\Om}(T)}+e^{-4\bar{C}_{\Om}(T)}}{2}f^2,
\end{eqnarray}
where $\bar{C}_\Om(T)\equiv\int_0^Tdt_1\int_0^{t_1}dt_2C_\Om(t_1,t_2)$. Here, $e^{-\bar{C}_{\Om}(T)}$ can be considered as a decay function, which is clearly in the range $(0, 1]$.

\begin{figure}[htbp]
\centering
\includegraphics[width=3.2in]{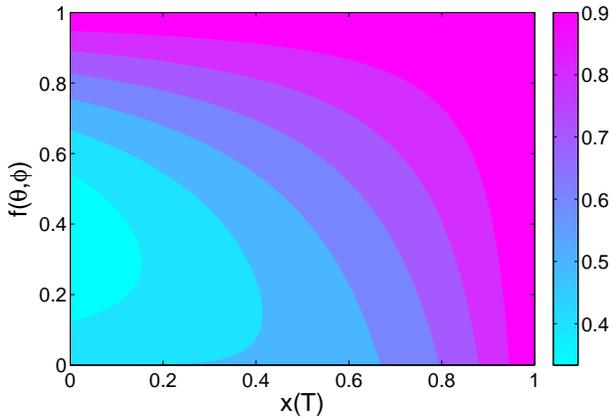}
\caption{(Color online) Landscape of the transformation fidelity $\mathcal{F}_\Om$ in the parameter space of decay function $x(T)\equiv e^{-\bar{C}_{\Om}(T)}$ and $f\equiv f(\theta,\phi)$.}\label{fig:Omega}
\end{figure}

Evidently, this transformation fidelity depends on the correlation function of the classical noise $\de_\Om$, the initial state characterized by $\alpha$, $\beta$, and $\eta$, and the control parameters $\theta$ and $\phi$. In Fig.~\ref{fig:Omega}, we provide a landscape of $\mathcal{F}_\Om$ plotted against the decay function $x(T)=e^{-\bar{C}_{\Om}(T)}$ and $f(\theta,\phi)$ according to Eq.~(\ref{fidOm}). Note that the impacts of the noise correlation function are already considered via $x(T)$ while those of the control phases $\theta$ and $\phi$ and input state parameters $\alpha$, $\beta$ and $\eta$ are included via $f(\theta,\phi)$. We thus have considered all possible regimes.

It is interesting to note that when $f(\theta,\phi)$ is close to unity, $\mathcal{F}_\Om$ is maintained at a high level for the whole range of the decay function $e^{-\bar{C}_{\Om}(T)}$, which depends on both $T$ (the cyclic period for constructing the logic subspace) and the form of the noise correlation function. For those particular combinations of quantum gates and input states, the gate operation is then robust against the classical noise even for a long runtime, noting that $x(T)$ often decreases with $T$ especially after coarse graining over the time domain. Similarly, when $x(T)$ is close to unity (larger than about $0.96$), corresponding to a short runtime, the transformation is found to be fault tolerant in the whole range of $f(\theta,\phi)$. Therefore, there are clearly ``sweet spots'' at $x(T)\simeq 1$ and $f(\theta,\phi)\simeq 1$.

A perfect ``sweet spot'' for this control problem (in view of quantum state engineering) emerges only when $f(\theta,\phi)=1$. From Eq.~(\ref{fidOm}), this gives rise to $\mathcal{F}_\Om=1$ irrespective of the existence of the stochastic fluctuations over the envelop function $\Om$. The condition for $f(\theta,\phi)=1$ to hold is $\cos(\phi+\eta)=1$ when $\alpha\beta\leq0$, or $\cos(\phi+\eta)=-1$ when $\alpha\beta\geq0$ and $\alpha=\pm\cos(\theta/2)$. Note that in the derivation, we apply the fact that $f(\theta,\phi)\leq1$. For example, if the input state is chosen as $\cos(3\pi/8)|0\ra+\sin(3\pi/8)|1\ra$, the Hadamard gate is always error free even if $\Om(t)$ is noisy.

\begin{figure}[htbp]
\centering
\includegraphics[width=3.2in]{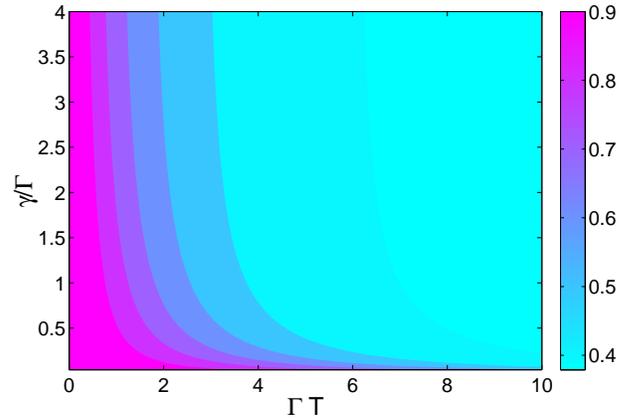}
\caption{(Color online) Average transformation fidelity $\bar{\mathcal{F}}_\Om$ under non-Markovian processing caused by $\de_\Om(t)$, as a function of dimensionless cyclic period $\Gamma T$ , and the memory parameter of noise $\gamma/\Gamma$. Here the correlation function is supposed to be $C_\Om(t,s)=\Gamma\gamma e^{-\ga|t-s|}/2$, so that the decay function $x=x(T)\equiv e^{-\bar{C}_\Om(T)}=\exp[-\Gamma(e^{-\ga T}+\ga T-1)/(2\ga)]$. When $\ga\rightarrow\infty$, $C_\Om(t,s)$ reduces to a delta function $\Ga\de(t-s)$ implying a purely Markovian noise and then $x$ reduces to the exponential decay function. On the other hand, for $\ga\rightarrow0$, the decay is strongly suppressed and the function $x(T)$ approaches unity.}\label{fig:aveOmega}
\end{figure}

An average fidelity $\bar{\mathcal{F}}_\Om$ over all input states is also studied. We adopt the parametrization $\alpha=\cos(\varphi/2)$ and $\beta=\sin(\varphi/2)$ and assume that $\varphi \in [0, \pi]$ and $\eta \in [0, 2\pi]$ follow uniform probability distributions. We then find that the average of $f(\theta, \phi)$ is $1/2$ and
\begin{equation}
\bar{\mathcal{F}}_\Om=\frac{3+4e^{-\bar{C}_{\Om}(T)}
+e^{-4\bar{C}_{\Om}(T)}}{8},
\end{equation}
which has a minimum of $3/8$. In Fig.~\ref{fig:aveOmega}, we plot the average fidelity assuming an Ornstein-Uhlenbeck noise with a correlation function
\begin{equation*}
C_\Om(t,s)=\frac{\Gamma\gamma}{2}e^{-\ga|t-s|},
\end{equation*}
where $\Ga$ is the correlation intensity of the noise and $\ga$ is the memory parameter and is inversely proportional to the memory retention time of the classical noise $\de_\Om$. In a strongly non-Markovian regime with $\ga/\Gamma\sim0.1$, the transformation fidelity can be maintained beyond $0.99$ for $T \alt 6/\Gamma$, a limit which is almost $12$ times as long as that in a near-Markovian case with $\ga/\Gamma\sim4$. Figure~\ref{fig:aveOmega} thus imposes an explicit demand on the runtime $T$ for HQC in the presence of the classical noise with different memory capabilities. As the correlation function of the noise approaches a delta function, the holonomic quantum gate runtime must become ever shorter.

We now calculate the minimal value of the fidelity for various initial states. From Eq.~(\ref{fidOm}), we get
\begin{eqnarray*}
\frac{\pa\mathcal{F}_{\Om}}{\pa f}&=&(3-4x+x^4)f-(x^4-2x+1), \\
\frac{\pa^2\mathcal{F}_{\Om}}{\pa f^2}&=&3-4x+x^4,
\end{eqnarray*}
where we have simplified the notation by writing $x\equiv e^{-\bar{C}_{\Om}(T)}$. Recall that $x\in(0,1]$, where the two bounds correspond to an infinitely large $T$ (or a strongly correlated noise) and a vanishing $T$ (or a Markovian noise) respectively. Consequently the second derivative $\pa^2\mathcal{F}_{\Om}/\pa f^2$ is always positive. Thus $\mathcal{F}_{\Om}$ attains its minimum when $\pa\mathcal{F}_{\Om}/\pa f=0$. Since the first term in $\pa\mathcal{F}_{\Om}/\pa f$ is positive, it may only vanish if $x \le  x_c\approx0.5437$, which is the only real root of $x^4-2x+1=0$ for $x\in[0,1)$. For $x\leq x_c$, the minimum fidelity occurs if $f(\theta,\phi)=(x^4-2x+1)/(3-4x+x^4)$. In contrast, for $x>x_c$, $\pa\mathcal{F}_{\Om}/\pa f \neq 0$. The minimum simply occurs at $f(\theta,\phi)=0$ that follows from Eq.~(\ref{fidOm}). This corresponds to the initial states satisfying $\alpha/\beta=\tan(\theta/2)$ for $\phi+\eta=2k\pi$, or $\alpha/\beta=-\tan(\theta/2)$ for $\phi+\eta=(2k+1)\pi$, with $k$ an integer.

\subsection{Fidelity under noisy control phases $\theta$ and $\phi$}\label{randomphase}

We now consider a noise-free envelop function $\Om(t)$. The holonomic quantum gate then possesses an exact cyclic time $T$. Fluctuations in $\theta$ or $\phi$ leave the system in the computational subspace spanned by the ground states $|0\ra$ and $|1\ra$ without invoking the excited state $|e\ra$. In the presence of random fluctuations associated with $d\theta/dt$, we have $\theta\rightarrow\theta'=\theta+\De_\theta(t)$, where $\De_\theta(t)=\int_0^tds\de_\theta(s)$. Consequently,
\begin{equation*}
\la\Psi(T)|\psi_{\theta}(T)\ra=\cos[\De_{\theta}(t)]
+2\alpha\beta\cos(\phi+\eta)\sin[\De_{\theta}(t)].
\end{equation*}
Inserting it into Eq.~(\ref{fidelity}), it is straightforward to show
\begin{equation}
\mathcal{F}_{\theta}=\frac{1+e^{-4\bar{C}_{\theta}(T)}}{2}
+4\alpha^2\beta^2\cos^2(\phi+\eta)\frac{1-e^{-4\bar{C}_{\theta}(T)}}{2},
\end{equation}
where $\bar{C}_{\theta}(T)\equiv\int_0^Tdt_1\int_0^{t_1}dt_2
\la\de_\theta(t_1)\de_\theta(t_2)\ra$. The ``sweet spot'' for this situation, i.e., $\mathcal{F}_{\theta}=1$, regardless of the existence of the noise, then emerges when $\alpha^2=1/2$ and $\phi+\eta=k\pi$, with $k$ an integer. In addition, the minimum fidelity occurs when $\alpha^2\beta^2\cos^2(\phi+\eta)=0$, in which $\mathcal{F}_{\theta}=1/2+\exp[-4\bar{C}_{\theta}(T)]/2$. Therefore, for a specific initial phase $\eta$ satisfying $\phi+\eta=(k+1/2)\pi$, the transformation fidelity is purely dependent on the correlation function of the noise $\de_{\theta}$, but independent of the population distribution $\alpha^2$ and $\beta^2$ of the initial state. On average over $\alpha$, $\beta$, and $\eta$, the fidelity turns out to be
\begin{equation}
\bar{\mathcal{F}}_{\theta}=\frac{5+3e^{-4\bar{C}_{\theta}(T)}}{8},
\end{equation}
which has a lower bound of $5/8$, larger than that for the case with a random envelop function $\Om'(t)$.

\begin{figure}[htbp]
\centering
\includegraphics[width=3.2in]{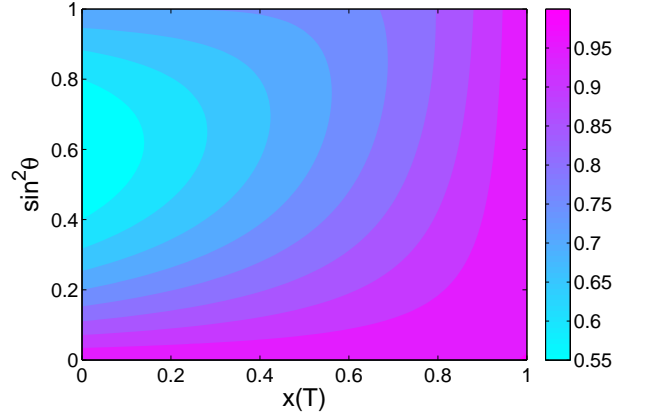}
\caption{(Color online) Landscape of average transformation fidelity $\bar{\mathcal{F}}_\phi$ in the parameter space of $x(T)\equiv e^{-\bar{C}_{\phi}(T)}$ and $\sin^2\theta$.}\label{fig:avephi}
\end{figure}

Similarly, in the presence of a random control phase $\phi\rightarrow\phi'=\phi+\De_\phi(t)$, where $\De_\phi(t)=\int_0^tds\de_\phi(s)$, we have
\begin{eqnarray*}
&& \la\Psi(T)|\psi_\phi(T)\ra=\sin^2\theta\left[\alpha^2e^{i\De_{\phi}(T)}
+\beta^2e^{-i\De_{\phi}(T)}\right]\\ &+&\cos^2\theta+ i\alpha\beta\sin(2\theta)[\sin(\phi'+\eta)-\sin(\phi+\eta)].
\end{eqnarray*}
After a tedious but straightforward derivation, the transformation fidelity is obtained as
\begin{eqnarray}\non
\mathcal{F}_{\phi}&=&\cos^4\theta+\sin^4\theta\left[1-
2\alpha^2\beta^2(1-e^{-4\bar{C}_{\phi}})\right] \\ \non &+&
\frac{\sin^2(2\theta)}{2}e^{-\bar{C}_{\phi}}+
\alpha^2\beta^2\sin^2(2\theta) \\ \non &\times&
\left[\sin^2\ti{\phi}(1-2e^{-\bar{C}_{\phi}})
+\frac{1-\cos(2\ti{\phi})e^{-4\bar{C}_{\phi}}}{2}\right]\\ \label{Fphi} &+&
\alpha\beta(\alpha^2-\beta^2)\sin^2\theta\sin(2\theta)
\cos\ti{\phi}(1-e^{-4\bar{C}_{\phi}}),
\end{eqnarray}
where $\bar{C}_{\phi}(T)\equiv\int_0^Tdt_1\int_0^{t_1}dt_2
\la\de_\phi(t_1)\de_\phi(t_2)\ra$ and $\ti{\phi}=\phi+\eta$.

Rather than locating the ``sweet spot" directly from Eq.~(\ref{Fphi}), it is more instructive to first average over $\alpha$, $\beta$ and $\eta$ and obtain
\begin{equation}
\bar{\mathcal{F}_{\phi}}=1-\frac{3\sin^2(2\theta)}{8}(1-e^{-\bar{C}_{\phi}})
-\frac{\sin^4\theta}{4}(1-e^{-4\bar{C}_{\phi}}).
\end{equation}
An interesting observation here is that in the presence of a noisy $\phi$, the average of the transformation fidelity depends on the particular value of $\theta$ in addition to the time-integrated noise correlation $\bar{C}_{\phi}$. The lower-bound of the average fidelity is found to be $1-3\sin^2(2\theta)/8-\sin^4\theta/4$. Therefore, the transformation fidelity depends on the particular type of the quantum gate determined by $\theta$. In Fig.~\ref{fig:avephi}, we show a general landscape of the average fidelity. We see that when $\sin^2\theta$ is sufficiently small, the average transformation fidelity shows a ``sweet spot'' regime. On the other hand, if the decay function $x(T)=e^{-\bar{C}_{\phi}(T)}$ is sufficiently large, equivalently if $T$ is sufficiently small or if the correlation function is in a strong non-Markovian regime, we still have an average fidelity close to unity for an arbitrary $\sin^2\theta$. In addition, it is found that when $\sin^2\theta=3/5$, $\bar{\mathcal{F}_{\phi}}$ arrives at its minimum value $11/20$, which is larger than that in the presence of a fluctuating envelop function $\Om'(t)$ and smaller than that in the presence of a fluctuating $\theta'$.

Considering averages over input states, values of the transformation fidelity minimized with respect to various gates and input states follow $\bar{\mathcal{F}}_{\Om}^{\rm min}<\bar{\mathcal{F}}_{\phi}^{\rm min}<\bar{\mathcal{F}}_{\theta}^{\rm min}$. In summary, the reliability of this nonadiabatic non-Abelian quantum gate is most susceptible to fluctuations occurring on the envelop function of Rabi frequency but is most resilient against that on the control phase $\theta$.

\section{Two-qubit gate}\label{discuss}

We have considered a direct and exact construction of a single-qubit nonadiabatic non-Abelian holonomic quantum gate by modulating two laser pulses interacting with a three-level atomic system (see Sec.~\ref{theory}). In contrast, an existing design~\cite{nanaholonomic,twoqubit} of a two-qubit gate is only approximately holonomic as a result of an adiabatic elimination under a restricted regime of the coupling strength between the laser and atoms. In the so-called S{\o}rensen-M{\o}lmer setting~\cite{twoqubit}, a pair of ions constitute two internal $\Lambda$-configuration three-level systems. The transitions $|e\ra\leftrightarrow|1\ra$ and $|e\ra\leftrightarrow|0\ra$ for these two ions are coupled by lasers with envelop functions of Rabi frequencies $\Om_1(t)$ and $\Om_0(t)$ and detunings $\pm\nu\pm\de$ and $\pm\nu\mp\de$, respectively, where $\nu$ is a phonon frequency and $\de$ is an additional detuning. The indirect interaction between the two ions is induced by the lasers. When the Lamb-Dicke parameter $\zeta$ satisfies $\zeta^2\ll1$, the effective Hamiltonian can be approximated as
\begin{eqnarray}
H_2&=&\frac{\zeta^2}{\de}\sqrt{\Om_0^4(t)+\Om_1^4(t)}H_2^{0}, \\ \non
H_2^{0}&=&\sin\frac{\theta}{2}e^{i\frac{\phi}{2}}|ee\ra\la00|
-\cos\frac{\theta}{2}e^{-i\frac{\phi}{2}}|ee\ra\la11|+h.c.,
\end{eqnarray}
where $\theta=2\tan^{-1}(\Om_0^2/\Om_1^2)$ and $\phi$ is the phase difference of two control pulses. Note now $\theta$ is no longer an independent physical parameter in contrast to its counterpart in the single-qubit protocols. In the protocol for two-qubit gate provided by Ref.~\cite{nanaholonomic}, the parameter $\theta$ is determined by the ratio of the amplitudes $\Om_0$ and $\Om_1$ of the pulse pair but not a phase under control as that for the single-qubit gate. To achieve a desired two-qubit gate, the ratio $\Om_0^2/\Om_1^2$ and the phase $\phi$ should be kept constant during each pulse pair. Meanwhile, $\Om_0$ and $\Om_1$ are constrained by the $\pi$ pulse criterion $\int_0^Tdt\frac{\zeta^2}{\de}\sqrt{\Om_0^4(t)+\Om_1^4(t)}=\pi$ to attain an effective evolution operator on the computational subspace spanned by $\{|00\ra, |01\ra, |10\ra, |11\ra\}$ forming a holonomic two-qubit gate
\begin{eqnarray}\non
\ti{U}_h&=&\cos\theta|00\ra\la00|-\cos\theta|11\ra\la11|+\sin\theta e^{-i\phi}|00\ra\la11| \\ &+& \sin\theta e^{i\phi}|11\ra\la00|+|01\ra\la10|
+|10\ra\la01|.
\end{eqnarray}
We parameterized the input state as $|\psi(0)\ra=\alpha|00\ra+\epsilon|01\ra+\eta|10\ra+\beta|11\ra$, where the four coefficients are real and follow the normalization condition. Then, we consider the classical noise perturbing the control phase difference $\phi$, which is a physically relevant parameter and independent from $\Om_0$ and $\Om_1$. Thus $\phi\rightarrow\phi'+\De_\phi(t)$, where $\De_\phi(t)=\int_0^tds\de_\phi(s)$. We study the decay function expressed by $x(T)=e^{-\bar{C}_\phi(T)}$, where the definition of $\bar{C}_\phi(T)$ in Sec.~\ref{randomphase} for the single-qubit case remains valid. The overlap between the output states of the ideal and nonideal unitary transformations is found to be
\begin{eqnarray*}
\la\Psi(T)|\psi_\phi(T)\ra&=&1-\alpha^2\sin^2\theta\left[1-e^{i\De_\phi(T)}\right]
\\ &-&\beta^2\sin^2\theta\left[1-e^{-i\De_\phi(T)}\right].
\end{eqnarray*}
Therefore, the transformation fidelity reads
\begin{eqnarray*}
\mathcal{F}_\phi^{(2)}&=&1-2\sin^2\theta[(\alpha^2+\beta^2)(1-x)]\\
&+&2\sin^4\theta[(\alpha^4+\beta^4)(1-x)+\alpha^2\beta^2(1+x^4-2x)],
\end{eqnarray*}
where $x=x(T)=e^{-\bar{C}_\phi(T)}$. Averaging over $\theta$, $\alpha$ and $\beta$, we find
\begin{equation}
\bar{\mathcal{F}}_\phi^{(2)}=1-\frac{93}{256}(1-x)+\frac{3}{64}(1+x^4-2x),
\end{equation}
which is a monotonic increasing function of $x$ in the range $(0,1)$ and tightly lower-bounded by $175/256\approx0.6836$. Thus in this particular situation, the minimum value of the average transformation fidelity for the two-qubit gate is even larger than that for the single-qubit gate.

\section{Discussion and conclusion}\label{conc}

In this work, for the nonadiabatic non-Abelian holonomic quantum single-qubit gate, we obtain general dependence of the transformation fidelity on the input (initial) states, the quantum gate parameters, and statistical properties of classical noises acting on these parameters. The gate parameters considered include the envelop function of Rabi frequency $\Om(t)$, and two control phases $\theta$ and $\phi$ of two driving laser pulses. The former one determines the runtime of the gate while the latter two determine the type of the gate. Values of the transformation fidelity averaged or minimized over all input states are also studied. The location of the high-fidelity regimes, i.e., the ``sweet spots'' implies that the nonadiabatic non-Abelian holonomic quantum gate is often robust against fluctuations in the control parameters. In the presence of the noise, we find that perfect ``sweet spot'' does exist under certain conditions, in particular in the case of nonideal parameter $\Om(t)$ or $\theta$. We also find a nonvanishing memory of the classical noise can relieve the requirement on the speed of a cyclic evolution of the logic subspace.

We then extend the analysis into a special two-qubit quantum gate leading to a universal set of quantum gates. It is interesting to find that the nonadiabatic non-Abelian holonomic two-qubit gate is more robust than the single-qubit gate against classical noise for the setups we have considered.

In conclusion, we have investigated nonadiabatic non-Abelian holonomic quantum gates for both single-qubit and two-qubit operations. By studying a unitary transformation fidelity, we clarify generic properties concerning the gate reliability, which are independent of the details of the classical noise correlation function. We compare the effect of classical noise from different sources. The analyses on the ``sweet spot'' and minimum values of the transformation fidelity are general and apply to Gaussian noise with arbitrary correlation functions. Our investigation provides a systematic estimation over the error of HQC caused by classical noise. It is expected to be useful for optimizing the performance of quantum gates for nonadiabatic non-Abelian holonomic quantum computing.

\section*{Acknowledgments}

We acknowledge grant support from the Basque Government (Grant No. IT472-10), the Spanish MICINN (Grant No. FIS2012-36673-C03-03), the National Science Foundation of China Grant No. 11575071, Science and Technology Development Program of Jilin Province of China (Grant No. 20150519021JH), and the HongKong GRF Grant No. 501213. We also acknowledge the helpful discussions with Dr. Junxiang Zhang and Dr. Fengdong Jia on the exclusive modulation over the parameters of Rabi frequency in experiments.

\end{document}